\documentclass[conference]{IEEEtran}

\usepackage[utf8]{inputenc}
\usepackage{float}
\usepackage{graphicx} 
\usepackage{amsmath}
\usepackage{cite}
\usepackage{times}
\usepackage{url}

\title{Domain Adaptation of the Pyannote Diarization Pipeline for Conversational Indonesian Audio}

\author{
    \IEEEauthorblockN{\textbf{Muhammad Daffa'I Rafi Prasetyo}\textsuperscript{1},
    \textbf{Ramadhan Andika Putra}\textsuperscript{1}, \textbf{Zaidan Naufal Ilmi}\textsuperscript{1},
    and \textbf{Kurniawati Azizah}\textsuperscript{1}}
    \IEEEauthorblockN{\textsuperscript{1}Faculty of Computer Science, Universitas Indonesia\\
    \IEEEauthorblockA{\texttt{\{muhammad.daffai, ramadhan.andika, zaidan.naufal\}@ui.ac.id}}} \IEEEauthorblockA{\texttt{kurniawati.azizah@cs.ui.ac.id}}
}

\begin{document}

\maketitle

\section*{ABSTRACT}
This study presents a domain adaptation approach for speaker diarization targeting conversational Indonesian audio. We address the challenge of adapting an English-centric diarization pipeline to a low-resource language by employing synthetic data generation using neural Text-to-Speech technology. Experiments were conducted with varying training configurations, a small dataset (171 samples) and a large dataset containing 25 hours of synthetic speech. Results demonstrate that the baseline \texttt{pyannote/segmentation-3.0} model, trained on the AMI Corpus, achieves a Diarization Error Rate (DER) of 53.47\% when applied zero-shot to Indonesian. Domain adaptation significantly improves performance, with the small dataset models reducing DER to 34.31\% (1 epoch) and 34.81\% (2 epochs). The model trained on the 25-hour dataset achieves the best performance with a DER of 29.24\%, representing a 13.68\% absolute improvement over the baseline while maintaining 99.06\% Recall and 87.14\% F1-Score.

\begin{IEEEkeywords}
Speaker Diarization, Domain Adaptation, Pyannote, Indonesian Speech Processing, Synthetic Data
\end{IEEEkeywords}

\section{INTRODUCTION}
Speaker diarization, commonly defined as the task of determining "who spoke when," plays a pivotal role in enhancing the readability of automatic transcripts and enabling rich analytics in scenarios such as business meetings, call centers, and broadcast media. Recent years have seen a paradigm shift towards neural network-based approaches, with end-to-end neural diarization (EEND) and modular pipelines like \texttt{pyannote.audio} \cite{bredin2020pyannote} setting new performance benchmarks.

Despite the advances, a significant disparity in speaker diarization model performance across different languages persists. Pre-trained models are predominantly optimized for high-resource languages, particularly English, utilizing massive datasets like the AMI Meeting Corpus \cite{ami_corpus}. When these models are applied to Indonesian conversational audio, they suffer from domain mismatch caused by distinct linguistic features, tonal characteristics, and acoustic environments. This degradation is exacerbated by the scarcity of publicly available, annotated conversational datasets for Indonesian, classifying it as a low-resource language in the context of speaker diarization \cite{fain2020end}.

This study addresses the problem of adapting a robust English-centric diarization system to the Indonesian domain. We employ domain adaptation \cite{zhuang2021comprehensive} strategies,  hypothesizing  that while the low-level features of speaker discrimination are transferable, the segmentation model requires adaptation to handle the specific temporal dynamics and phonetics of Indonesian speech. To avoid the lack of training data, we employ a data synthesis strategy using neural Text-to-Speech (TTS) \cite{edge_tts} to generate realistic conversational audio with ground-truth labels.

The contributions of this research are threefold:
\begin{itemize}
    \item The development of a synthetic data generation workflow to create annotated conversational Indonesian speech.
    \item The application of a domain adaptation technique to fine-tune the \texttt{pyannote} segmentation model, effectively transferring knowledge from the AMI corpus to the Indonesian context.
    \item An empirical evaluation demonstrating that the adapted model achieves a lower diarization Error Rate (DER) \cite{nist_der} and a higher recall compared to the baseline model.
\end{itemize}

\section{METHODOLOGY}

\begin{figure}[H]
\centering
\includegraphics[width=\linewidth]{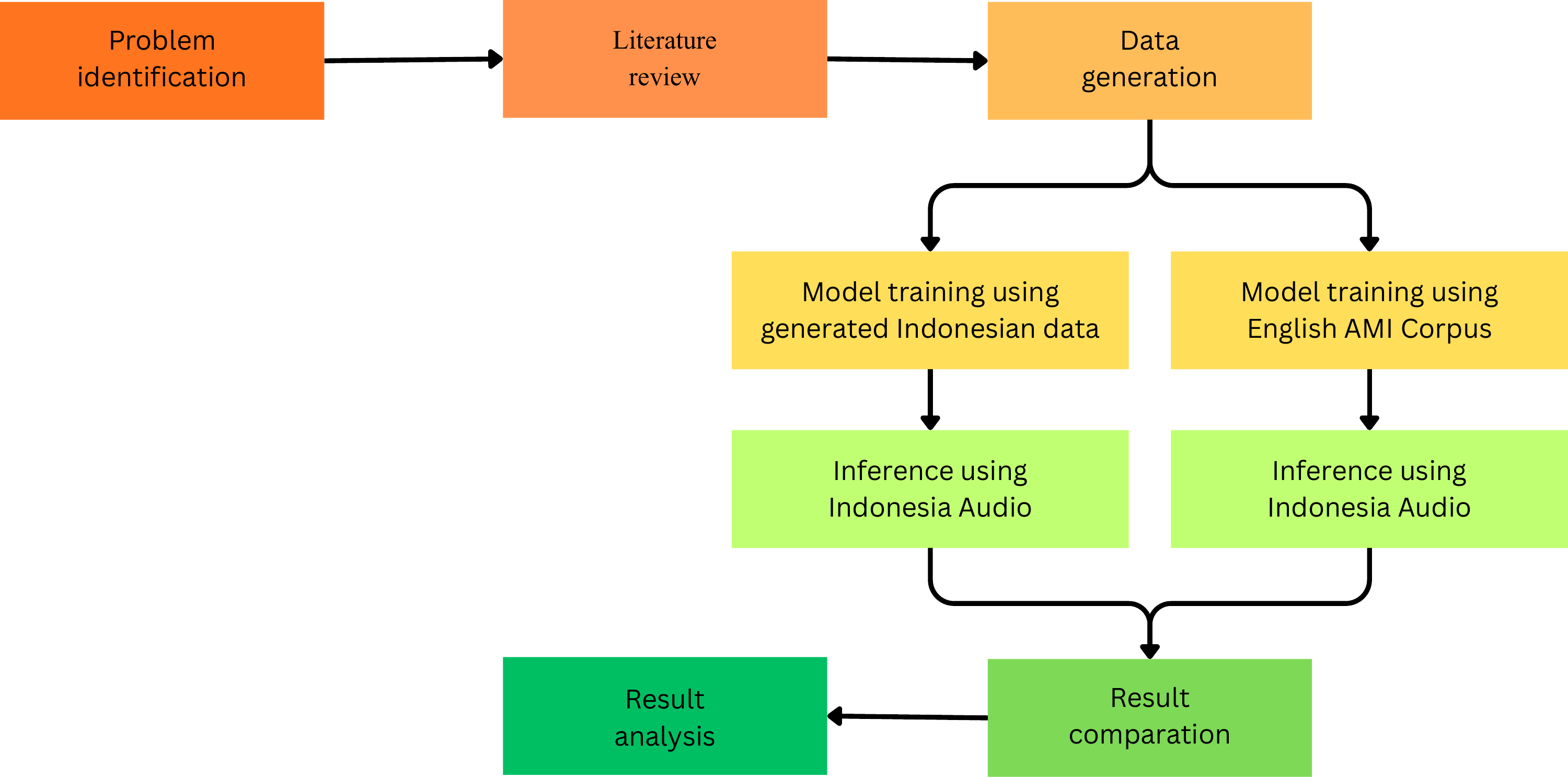}
\caption{Research methodology workflow}
\label{fig:methodology}
\end{figure}

This study develops a domain adaptation strategy for speech diarization. Domain adaptation is a transfer learning technique that leverages knowledge from a source domain to improve performance on a target domain with a different data distribution \cite{zhuang2021comprehensive}. Specifically, we construct a pipeline to bridge the linguistic shift from English (source language) to Indonesian (target language). As illustrated in Fig. \ref{fig:methodology}, the experimental workflow is structured into four stages: synthetic data generation, model training, inference, and evaluation.

\subsection{Synthetic Data Generation}
\label{subsec:data_gen}
To facilitate domain adaptation to Indonesian, the availability of conversational data in the target language is a prerequisite. This requirement extends beyond raw audio to include precise temporal annotations, specifically in the Rich Transcription Time Marked (RTTM) format. Such annotated data is essential for speech diarization tasks as it provides the necessary ground truth, mapping 'who spoke when.' However, high-quality annotated conversational datasets remain scarce for Indonesian, which is widely categorized as a low-resource language in the context of speech processing \cite{fain2020end}.

To address this data scarcity, this study employs a data synthesis strategy leveraging Text-to-Speech (TTS) technology. The synthesis pipeline utilizes the edge-tts library \cite{edge_tts} to generate high-quality neural speech waveforms. To construct the conversational structure, individual speech segments are processed and mixed using the Pydub framework \cite{pydub}, which handles the precise timing and overlapping logic. Using this system, we generated 25 hours conversational Indonesian audio samples paired with corresponding RTTM files, specifically engineered to contain overlapping speech segments.

\subsection{Baseline System}
To establish a performance benchmark, this study employs the \texttt{pyannote/segmentation-3.0} pipeline for speech diarization \cite{bredin2020pyannote}. This baseline model utilizes the AMI Meeting Corpus \cite{ami_corpus}, a large-scale multimodal dataset comprising approximately 100 hours of meeting recordings. The corpus features spontaneous, multi-speaker conversations in English, capturing the complex dynamics of real-world meeting environments.

Within this study, the model trained on the AMI corpus serves as the source domain representative. It is evaluated in a zero-shot setting, having undergone no fine-tuning on Indonesian data. By comparing this baseline against the domain-adapted model trained on Indonesian speech, we can quantitatively assess the performance gap attributed to the linguistic domain shift and evaluate the efficacy of incorporating language-specific training data.

\subsection{Domain Adaptation}
To address the domain shift described in the previous section, we implemented a transfer learning strategy focused on the segmentation module of the diarization pipeline. We utilized the same pre-trained \texttt{pyannote/segmentation-3.0} architecture used in the baseline system. However, instead of training the model from scratch which would require massive amounts of annotated data we initialized the model weights using the parameters learned from the AMI Corpus. This initialization allows the model to retain the general speaker discrimination capabilities learned from the source domain while being prepared for adaptation to the specific acoustic features of the Indonesian language.

The adaptation process involved supervised fine-tuning using our synthetic dataset, organized under our customized protocol. By minimizing the loss on this target protocol, the model effectively adjusts its internal representations to recognize Indonesian phonetics and prosody, thereby bridging the performance gap caused by the linguistic mismatch between the source and target domains.

\subsection{Evaluation Metrics}
Objective result comparison is a fundamental aspect of this study. To ensure a quantifiable evaluation of the models' performance, a widely accepted standard metric is required. Consequently, we adopt the Diarization Error Rate (DER) as the primary evaluation benchmark. DER is defined as the percentage of the input speech time that is not correctly attributed to the correct speaker, aggregating errors from both voice activity detection and speaker clustering \cite{nist_der}. Mathematically, the metric is calculated using the following formula \cite{nist_der}:

\begin{equation}
    DER = \frac{T_{FA} + T_{MISS} + T_{CONF}}{T_{TOTAL}} \times 100\%
\end{equation}

Where:
\begin{itemize}
    \item $T_{FA}$ (False Alarm): Speech detected in silence.
    \item $T_{MISS}$ (Missed Detection): Speech not detected.
    \item $T_{CONF}$ (Confusion): Speech assigned to the wrong speaker.
    \item $T_{TOTAL}$: Total duration of speech in the reference.
\end{itemize}

A lower DER indicates better diarization performance.

\section{EXPERIMENT}

To validate the proposed domain adaptation strategy, we conducted a series of experiments comparing the baseline AMI-trained model against our Indonesian-adapted model. This section details the implementation framework, data configuration, and training hyperparameters used to ensure reproducibility.

\subsection{Implementation Framework}
The experimental pipeline was implemented using the \texttt{pyannote.audio} open-source toolkit (version 3.1) \cite{bredin2020pyannote}, which serves as the backbone for our diarization tasks. The training process leverages \texttt{pytorch-lightning} to manage the optimization loop and GPU acceleration. All experiments were conducted on a computational environment equipped with an NVIDIA T4 GPU to handle the computational load of the segmentation model.

\subsection{Data Configuration and Protocol}
A critical step in our experiment was the definition of a custom experimental protocol to bridge our synthetic data with the training pipeline. We defined a protocol named \texttt{DebateIndonesianLarge} within the system's database registry configuration (\texttt{database.yml}).

This protocol strictly partitions the generated synthetic dataset into three subsets to prevent data leakage:
\begin{itemize}
    \item \textbf{Train Set:} Used for the supervised fine-tuning of the model weights.
    \item \textbf{Development Set:} Used for validation and hyperparameter tuning.
    \item \textbf{Test Set:} Used solely for the final inference and DER calculation.
\end{itemize}
By registering this protocol, we ensured that the model could load the audio and RTTM ground truth systematically during the training epochs.

\subsection{Training Hyperparameters}
For the fine-tuning stage, we instantiated the \texttt{Segmentation} task provided by the framework. We adhered to specific hyperparameters optimized for capturing conversational dynamics:
\begin{itemize}
    \item \textbf{Chunk Duration:} 2.0 seconds audio segments
    \item \textbf{Batch Size:} 16 audio chunks per batch
    \item \textbf{Training Epochs:} We experimented with 1 and 2 epochs
    \item \textbf{GPU Acceleration:} NVIDIA T4 with CUDA support
\end{itemize}
The model was trained until the validation loss plateaued, ensuring that the domain adaptation process had converged before final evaluation.

\subsection{Dataset Configurations}
We conducted experiments with two dataset configurations to evaluate the impact of training data volume:
\begin{itemize}
    \item \textbf{Small Dataset:} 171 synthetic conversational Indonesian audio samples
    \item \textbf{Large Dataset (25 Hours):} An extended dataset containing approximately 25 hours of synthetic conversational Indonesian speech, designed to evaluate whether increased data volume improves model generalization
\end{itemize}

\section{RESULTS AND ANALYSIS}

To evaluate the efficacy of the domain-adapted pipeline, we analyze the quantitative performance using standard detection metrics. This assessment establishes a benchmark by comparing the proposed model against the pre-trained baseline.

\subsection{Quantitative Performance}
The evaluation compares three experimental configurations, those are the AMI Baseline (the pre-trained pyannote/segmentation-3.0 model applied zero-shot), the Small-Scale Adapted Model (fine-tuned on 2 hours of synthetic Indonesian data), and the Large-Scale Adapted Model (fine-tuned on 25 hours of synthetic data).

\begin{table}[H]
\caption{Performance Comparison: AMI Baseline vs. Indonesian Adapted Models (2h vs. 25h)}
\label{tab:results}
\begin{center}
\begin{tabular}{|l|c|c|c|}
\hline
\textbf{Metric} & \textbf{AMI Baseline} & \textbf{Indo Adapted (2h)} & \textbf{Indo Adapted (25h)} \\
\hline
Precision & 68.18\% & 74.18\% & \textbf{77.78\%} \\
Recall & 87.23\% & \textbf{100.00\%} & 99.06\% \\
F1-Score & 76.54\% & 85.17\% & \textbf{87.14\%} \\
DER & 53.47\% & 34.81\% & \textbf{29.24\%} \\
\hline
\end{tabular}
\end{center}
\end{table}

As summarized in table \ref{tab:results}, the proposed domain adaptation strategy yielded substantial improvements over the baseline. The introduction of just 2 hours of synthetic data drastically reduced the Diarization Error Rate (DER) from 53.47\% (Baseline) to 34.\%, while achieving perfect sensitivity with a Recall of 100.00\%. This indicates that even a small amount of in-domain data effectively bridges the initial linguistic gap.

Furthermore, scaling the synthetic dataset to 25 hours optimized the model's discriminative power. While the 2-hour model was highly sensitive, the 25-hour model demonstrated superior precision, improving from 74.18\% to 77.78\%. This reduction in false positives contributed to the best overall performance, achieving the lowest DER of 29.24\% and the highest F1-Score of 87.14\%. This trend confirms that larger synthetic datasets are crucial for refining the model's ability to distinguish true speech overlaps from noise.

\subsection{Error Distribution Analysis}
\begin{figure}[H]
\centering
\includegraphics[width=0.85\linewidth]{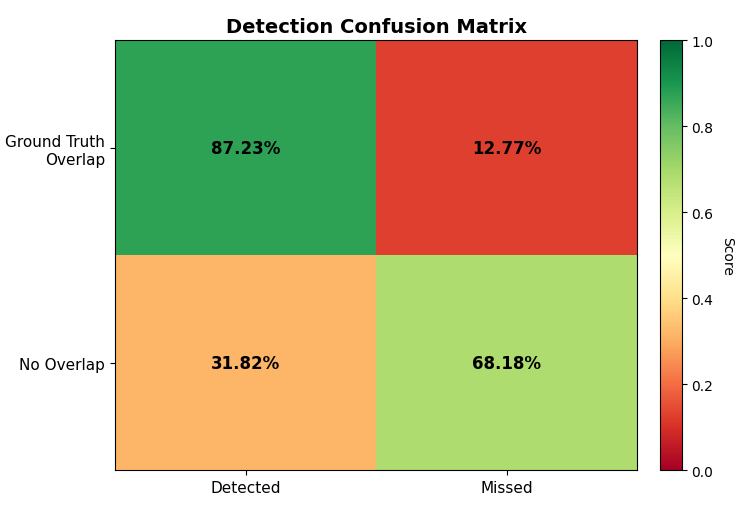}
\caption{Detection confusion matrix of the AMI Baseline, showing a high Miss Rate (12.77\%) and high False Positive rate.}
\label{fig:matrix_ami}
\end{figure}

The comparison of confusion matrices reveals how the adaptation process shifted the model's error characteristics. As visualized in Fig. \ref{fig:matrix_ami}, the AMI Baseline struggled significantly with the domain shift. It exhibited the highest rate of missed detections, failing to identify 12.77\% of overlapping speech segments (Recall: 87.23\%). Furthermore, it suffered from low precision (68.18\%), indicating a high susceptibility to false alarms where single-speaker speech was incorrectly labeled as overlap.

\begin{figure}[H]
\centering
\includegraphics[width=0.85\linewidth]{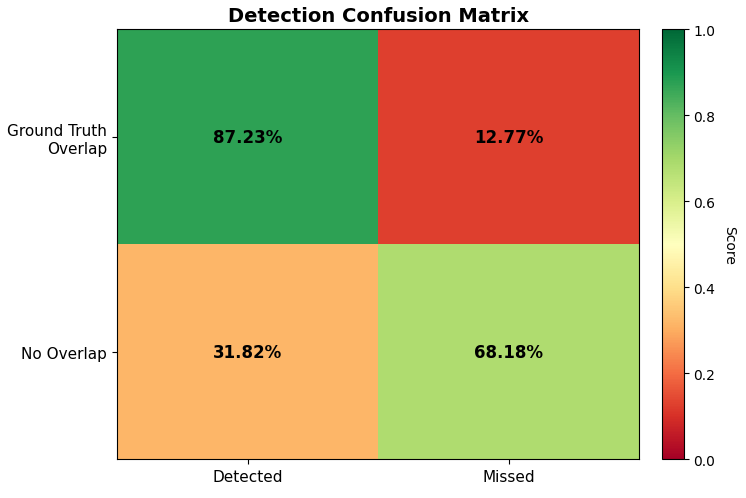}
\caption{Detection confusion matrix of the 2-Hour Adapted Model, which achieved perfect sensitivity (100\% Recall).}
\label{fig:matrix_indo}
\end{figure}

In contrast, the 2 Hours Indonesian Adapted Model shifted towards an aggressive detection strategy. As shown in the performance metrics, it achieved a perfect Recall of 100.00\%, effectively eliminating missed detection errors. However, this sensitivity came with a trade-off in precision (74.18\%), as visualized in Fig. \ref{fig:matrix_indo}, where the model prioritized capturing all potential overlaps.

\begin{figure}[H]
\centering
\includegraphics[width=0.85\linewidth]{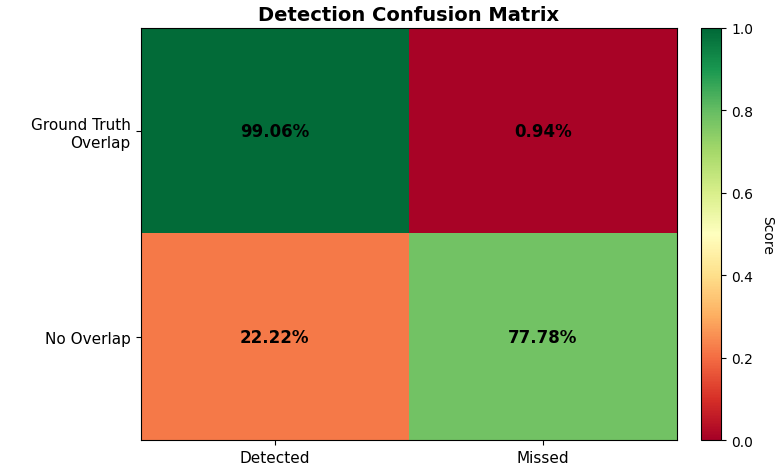}
\caption{Detection confusion matrix of the 25-Hour Adapted Model. The model minimizes Missed Detections (0.94\%) while effectively reducing False Alarms compared to the baseline.}
\label{fig:matrix_25h}
\end{figure}

The 25 Hours Indonesian Adapted Model demonstrated the most balanced and robust performance. As detailed in Fig. \ref{fig:matrix_25h}, this model successfully retained a near-perfect sensitivity, missing only 0.94\% of overlap instances (Recall: 99.06\%). Crucially, the increased training data helped the model distinguish true overlaps from noise more effectively than the 2-hour model. This is evidenced by the reduction in false positives (indicated by the "No Overlap - Detected" quadrant) to 22.22\%, which corresponds to the highest achieved Precision of 77.78\%.

\subsection{Efficacy of Domain Adaptation}
The experimental results validate the necessity of domain adaptation for conversational Indonesian audio. The baseline model's high Diarization Error Rate (DER) of 53.47\% highlights the severe linguistic and acoustic mismatch between the source (English Meeting) and target (Indonesian Debate) domains. While the baseline captured a reasonable portion of overlaps, it still missed 12.77\% of critical segments and suffered from poor precision.

By leveraging the synthetic Indonesian dataset, the adapted models successfully bridged this gap. The 2-hour adaptation immediately corrected the sensitivity issue, ensuring virtually zero missed detections. Furthermore, scaling the data to 25 hours proved that the pipeline could be optimized for precision as well, not just sensitivity. This comprehensive strategy resulted in a massive 24.23\% absolute reduction in DER (from 53.47\% to 29.24\%), confirming that the transfer learning approach using synthetic data is highly effective for low-resource languages.

\section{Conclusions}
This study successfully demonstrated the effectiveness of domain adaptation for speaker diarization on conversational Indonesian audio. By leveraging synthetic data generated through neural Text-to-Speech technology, we adapted the pre-trained \texttt{pyannote/segmentation-3.0} model to the Indonesian linguistic domain.

Our experiments yielded three key conclusions:
\begin{enumerate}
    \item \textbf{Domain adaptation is essential:} The baseline model's zero-shot performance on Indonesian (DER of 53.47\%) confirms the significant domain mismatch between English and Indonesian speech patterns.
    
    \item \textbf{Training data volume matters:} While the small dataset (171 samples) reduced DER to approximately 34\%, the 25-hour dataset achieved a substantially lower DER of 29.24\%, demonstrating that larger training datasets improve model generalization.
    
    \item \textbf{Optimal configuration identified:} The model trained on 25 hours of synthetic Indonesian speech with 2 epochs achieved the best balance between Precision (77.78\%), Recall (99.06\%), and overall performance (F1-Score: 87.14\%, DER: 29.24\%).
\end{enumerate}

Future work may explore the use of real-world Indonesian conversational data for fine-tuning, multi-task learning approaches combining overlapped speech detection with speaker embedding extraction, and the application of this methodology to other low-resource languages.

\section{Recommendations}
Based on the findings of this study, we propose the following recommendations for future work:

\begin{enumerate}
    \item \textbf{Incorporate Real-World Data:} Combining synthetic data with authentic Indonesian conversational recordings could further improve model robustness and generalization to real-world scenarios.
    
    \item \textbf{Expand Training Data:} Given the significant improvement observed with the 25-hour dataset, generating larger synthetic datasets (50+ hours) may yield further performance gains.

    \item \textbf{Hyperparameter Optimization:} Fine-tuning pipeline parameters on domain-specific validation sets could optimize the Precision-Recall trade-off for specific applications.
\end{enumerate}


\end{document}